\documentclass[12pt]{article}
\usepackage{epsfig}
\usepackage{latexsym}
\usepackage{graphics}
\begin{document}
\bibliographystyle{unsrt}

\begin{titlepage}
\begin{center}
\hfill CERN-TH/98-205 \\
\vspace{0.2cm}
\hfill RI-7-98 \\
\vspace{2cm}
{\Large\bf On Least Action D-Branes}\\
\vspace{1cm}
{\large Shmuel Elitzur}${}^{a,}$\footnote{e-mail address: elitzur@vms.huji.ac.il},
{\large Eliezer Rabinovici}${}^{a,b,}$\footnote{e-mail address: eliezer@vms.huji.ac.il} and 
{\large Gor Sarkissian}${}^{a,}$\footnote{e-mail address: gor@vms.huji.ac.il} \\
\vspace{0.5cm}
${}^a${\em Racah Institute of
Physics, The Hebrew University\\ Jerusalem 91904, Israel} \\
\vspace{0.5cm}
${}^b${\em CERN, 1211 Geneva 23, Switzerland}
\end{center}
\vspace{1cm}
\centerline{{\bf{Abstract}}}
We discuss the effect of relevant boundary terms on the nature of branes.
This is done for toroidal and orbifold compactifications of the bosonic string.
Using the relative minimalization of the boundary entropy as a guiding 
principle, we uncover the more stable boundary conditions at different
regions of moduli
space. In some cases, Neumann boundary conditions dominate for small radii
while Dirichlet boundary conditions dominate for large radii. The $c=1$ and
$c=2$ moduli spaces are studied in some detail.  The antisymmetric background
field $B$ is found to have a more limited
role in the case of Dirichlet boundary conditions. This is due to some
topological considerations. The results are subjected to $T$-duality tests
and the  special role of the points in moduli space fixed under $T$-duality
is explained from least-action considerations.

\vspace{3cm}
CERN-TH/98-205  \\

June 1998
\vfill
\end{titlepage}
\newpage

\section{Introduction}
String theory has perturbatively a large moduli space. Various ideas
have been put forward on how to remove at least part of this vast
degeneracy. One of these ideas called for checking the infrared stability
of all ground states \cite{POL}.

A point in moduli space whose spectrum contains operators
that are eventually relevant in the infrared seems to be unstable and
ceases to be itself part of the moduli space. In space-time language this
corresponds to the emergence of unstable directions in the effective low-energy
potential. It is less clear how this infrared instability is actually
resolved and
which vacuum replaces the unstable one. On the one hand, string theories
always have a vanishing overall Virasoro central charge $c$, while on the
 other
hand the central charge in the perturbed unitary sector decreases \cite{ZAM}.
Such issues can be addressed in the framework of so-called non-critical
string backgrounds \cite{EGR}.

Recently, new sectors of the moduli space were uncovered; they contain
different
brane configurations, some of them related by $T$-duality. One can reexamine
in this setting the issues of infrared stability, as well as the fate of the
``false" backgrounds. One can also consider the outcome of the addition of some
infrared relevant boundary perturbations. Such perturbations have the
feature that they leave unchanged the total bulk central charge. It will
turn out that they may change the $p$ dimensionality of the $Dp$
backgrounds.  This occurs since, at some points in moduli space, boundary
conditions are forced to change
from Dirichlet to Neumann ones and vice versa, see for example \cite{W}. The direction of flow
is determined by the value of a boundary entropy $g$ defined in \cite{AL}.
It was shown in \cite{AL2} that
as the system flows from one conformal theory to the other, in the presence of
a relevant boundary operator, the value of $g$ decreases.
This was exhibited to first order in conformal
perturbation theory suggesting a so called $g$ theorem similar to the $c$
theorem \cite{ZAM}. \\
This was studied in the presence of a Sine-Gordon boundary perturbation
\cite{FSW}; it was also pointed out \cite{CK,SCH} that, for the case of a
string with
Dirichlet boundary conditions, the boundary entropy can also be recognized as
the target space $Dp$ world-volume effective-action density.
In this paper we will identify the appropriate $Dp$ backgrounds for the
bosonic string theory. This will be done in detail for the cases of $c=1$
and $c=2$ and will be studied also for general toroidal compactifications.
In the absence of torsion background fields, Neumann boundary conditions will
be found to dominate at the lower values
of the target space volume, while Dirichlet boundary conditions dominate
when the appropriate target-space radii are large. The fixed points under
duality play the role of  boundaries in moduli space between the different
boundary conditions.

The paper is organized as follows:

In section $2$ we review the definition of the boundary entropy and obtain its
values for the full $c=1$ moduli space. This includes the  compactification
of one coordinate on a circle of radius $R$, an orbifold of radius $R$ as well as
the value at  the Ginsparg points \cite{G}. This is done in the presence of
a Wilson line. Special attention is given to the relation between the fixed
points in the moduli space and placements of the Wilson line.

We verify that the values of $g$ satisfy known properties of the $c=1$
moduli space such as $T$-duality and the coincidence between a string
moving in the presence of a circle or an orbifold for a special value
of their respective radii. A rather general derivation of $g$ is presented
for an orbifold compactification resulting from modding out the target
manifold by some discrete finite symmetry group $G$, of order $n_G$.
The dominant boundary conditions at each point in moduli space are
identified and the uncovered structure is motivated by energy considerations.

In section $3$ these issues are discussed for more general toroidal
compactifications. It is found that $g$ depends only on the components
of the background torsion field $B$ parallel to the $Dp$-brane.
 The topological reasons for that
are discussed. In section $4$ the results are subjected to $T$-duality
checks. In section $5$ the system with mixed Neumann and Dirichlet
boundary conditions is studied. The fixed points of $T$-duality are
shown to play the role of boundary points in moduli space. The structure
is exemplified by a detailed description of the dominant boundary conditions
in the case of the $c=2$ moduli space.

\section{The boundary entropy for c=1 systems}

The moduli space of $c=1$ compactifications includes a string propagating on a circle with
radius $R$,  an orbifold of radius $R$ \cite{EGRS} and isolated Ginsparg points. In this
section we consider an open bosonic string moving on these backgrounds and calculate
 the boundary entropy allowing also for Chan-Paton factors and for
the presence of a boundary interaction term,  the Wilson line,
 for both Dirichlet and Neumann boundary conditions.
We first recall the manner in which the boundary entropy is defined in the
general case and the way in which it is calculated.

\subsection{Definition of boundary entropy}
Let us consider a conformal field theory on the $\sigma-\tau$ strip, $0\leq
\sigma \leq \pi$, periodic in the $\tau$-direction
 with a period $T$.
The manifold  is an annulus with the modular parameter \mbox{$q\equiv \exp(-2\pi iT)$}.
Given certain boundary conditions on the boundaries of the annulus, labelled 
$\alpha$ and $\beta$,
 the partition function is:
\begin{equation}
\label{LC}
Z_{\alpha\beta}=\mathrm{Tr}\exp(-2\pi iT H_{\alpha\beta}) \ ,
\end{equation}
where $H_{\alpha\beta}$ is the Hamiltonian corresponding to these boundary conditions.
 This is the open-string channel.
 One may also calculate the partition function using the Hamiltonian acting in 
the $\sigma$-direction \cite{JC}.
This will be the Hamiltonian $H^{(P)}$ for the cylinder, which is related by the exponential mapping 
$\zeta=\exp(-i(t+i\sigma))$ to the Virasoro generators in the whole $\zeta$-plane by 
\mbox{$H^{(P)}=L_{0}^{(P)}+\overline{L_{0}}^{(P)}-c/12$}, where we have used the superscript to stress 
that they are not the same as the generators of the boundary Virasoro algebra.
 It was shown in \cite{JC}, that to every boundary condition $\alpha$, there
 corresponds
a particular boundary state $|\alpha\rangle$ in the Hilbert space of the closed strings; this
 enables us to compute the partition function by the following formula:

\begin{equation}
\label{TC}
Z_{\alpha\beta}=\langle\alpha|\exp(-\pi iH^{(P)}/T)|\beta\rangle =
\langle\alpha|(\tilde{q}^{1/2})^{L_{0}^{(P)}+\overline{L_{0}}^{(P)}-c/12}|\beta\rangle\
,
\end{equation}

where  \mbox{$\tilde{q}\equiv e^{-2\pi i/T}$}.

This is the closed-string tree channel. 

The boundary entropy   for each
boundary is defined by \cite{AL}:
\begin{equation}
  g_{\alpha}=\langle0|\alpha\rangle \ .
\end{equation}
The phases of $|0\rangle$ and $|\alpha\rangle$ can be chosen such that
 $\langle0|\alpha\rangle$ is real and positive for all boundary states $|\alpha\rangle$.
In the path integrand language,  $g_{\alpha}$ is the value of the disc diagramm
satisfying $\alpha$ type boundary condition.
 It was also shown in  \cite{AL2} that,  at
least in conformal perturbation theory, the value of $g$ always decreases with
the flow of the renormalization group.

The equality of (\ref{LC}) and  (\ref{TC}) provides a convenient way to
calculate $g$, as shown in the following.

\subsection{Boundary entropy for the $c=1$ string compactified on a circle}

\subsubsection{Neumann boundary conditions}
The action describing the bosonic $d=1$ string with the Wilson line boundary
 interaction is \cite{GR}:
\begin{equation}
\label{WILACT}
S=\frac{1}{2\pi}\int_{0}^{\pi}d\sigma\int
d\tau\partial_{\alpha}X\partial^{\alpha}X+\sum_{B}\frac{iy_{B}}{\pi}\int_{B}dX
\ ,
\end{equation}
where $B$ labels boundaries and $y_{B}$ are the constant modes of the $U(1)$
 gauge potential coupling to the
boundaries (and are periodic, with periods $\pi/R$).
We assume that the boundaries carry also Chan-Paton factors whose index 
we choose to take two values, $1$ and $2$. Thus at the enhanced symmetry point we have a $U(2)$
gauge symmetry,  which is generically broken down to $U(1)\times U(1)$ by the Wilson
line.

Here we consider a world-sheet with two boundaries, the annulus diagram. 

In order to find the boundary entropy, the theory should be compared
 in two channels: the closed-string tree channel and the open-string loop channel.

In the closed-string channel the first task is to find the boundary states
$|N_{i}\rangle$, with Chan-Paton factor $i$, which are found by imposing the
corresponding boundary conditions.
The boundary is located at $\tau=0$ and one has the
usual condition of vanishing momentum flow:
\begin{equation}
\partial_{\tau}X(\sigma,0)=P(\sigma,0)=0 \ .
\end{equation}
Inserting the mode expansion:
\begin{equation}
X^{i}(\sigma ,\tau )  =  x+2wR\sigma+ \frac{p\tau}{R} 
 + \frac{i}{2} \sum_{n \not= 0}\frac{1}{n} [\alpha_{n}e^{-2in(\tau -\sigma)}+
\tilde{\alpha}_{n}e^{-2in(\tau +\sigma)}] \ ,
\end{equation}
where $p$ and $w$ are correspondingly integer momenta and winding numbers,
 we get:
\begin{equation}
p=0,\hspace{1cm} \alpha_{n}=-\tilde{\alpha}_{-n}\ .
\end{equation}
 Taking into account the  properties of coherent state and
the $U(1)$ modes $y_{i}$ we get
for $|N_{i}\rangle$:
\begin{equation}
\label{NEU}
|N_{i}\rangle=g_{N_{i}}\sum_{w}e^{-2iy_{i}wR}\exp\left(\sum_{n>0}-\frac{\alpha_{-n}\tilde{\alpha}_{-n}}{n}\right)|0,w\rangle
\ ,
\end{equation}
where $w$ is the integer winding number.
We see that the normalization factor $g_{N_{i}}=\langle0|N_{i}\rangle$  gives
us the boundary entropy.
Inserting the expression for $|N_{i}\rangle$ and the closed string Hamiltonian
\begin{equation}
H=\frac{p^{2}}{4R^{2}}+w^{2}R^{2}+N+\tilde{N}-\frac{1}{12}
\end{equation}
 in (\ref{TC}), we obtain for the partition function in the closed string  channel:
\begin{eqnarray}
\label{P1}
Z_{i}=g_{N_{i}}^{2}\langle N_{i}|\exp\left(\frac{-i\pi
w^{2}R^{2}}{T}\right)\exp\left[\frac{-i\pi}{T}(N+\tilde{N}-\frac{1}{12})\right]
|N_{i}\rangle=  \nonumber \\
\frac{g_{N_{i}}^{2}}{\eta(\tilde{q})}\sum_{w}\exp\left(\frac{-i\pi
w^{2}R^{2}}{T}\right)
=\frac{g_{N_{i}}^{2}}{\eta(\tilde{q})}\theta_{3}\left(\frac{-R^{2}}{T},0\right)
\ ,
\end{eqnarray}
where \( \eta(q)=q^{1/24}\prod_{n=1}^{\infty}(1-q^{n}) \) is the Dedekind function, 
and  \( \theta_{3}(\tau,z)=\sum_{n=-\infty}^{\infty}\exp(i\pi n^{2}\tau+2inz) \) 
is the third theta function with the modular parameter $\tau$.
To calculate $g_{N_{i}}$ one turns to the open string loop channel.
 The  Hamiltonian should be computed with a given boundary condition.
First consider the mode expansion for $X$. The mode expansion of
the solution of the equation of motion is :
\begin{equation}
X=x+\frac{p\tau}{R}+i\sum_{n \not=0}\frac{1}{n}\alpha_{n}
 \cos (n\sigma) \exp(-in\tau)\ ,
\end{equation}
where $p$ is an integer.
 Inserting this in the open-string Hamiltonian, we obtain:
\begin{equation}
H=\frac{p^{2}}{2R^{2}}+N-\frac{1}{12}\ .
\end{equation}
 The partition function in this
channel is :
\begin{equation}
\label{P2}
Z=\frac{1}{\eta(q)}\sum_{p}\exp\left(\frac{-i\pi Tp^{2}}{R^{2}}\right)
=\frac{1}{\eta(q)}\theta_{3}\left(-\frac{T}{R^{2}},0 \right)\ .
\end{equation}
 Equating (\ref{P1}) and (\ref{P2})  and using the properties of modular transformations:
\begin{eqnarray}
\label{M}
\theta_{3}\left(\frac{1}{\tau},z\right)=\tau^{1/2}e^{i\tau z^{2}/\pi}\theta_{3}(\tau,\tau z) \\
\eta(\tilde{q})=(-T)^{1/2}\eta(q)\ ,
\end{eqnarray}
we obtain:
\begin{equation}
\label{entn}
g_{N_{i}}^{2}=R \ ,
\end{equation}
independent of the Wilson line parameter $y_{i}$.
It can be explained by elaborating the boundary interaction in (\ref{WILACT})
\cite{GR}:
\begin{equation}
\sum_{B}\frac{iy_{B}}{\pi}\int_{B}dX=\sum_{B}2iy_{B}w_{B}R\ ,
\end{equation}
where $w_{B}$ is the winding number of the boundary.
Since the boundary of a disc diagramm cannot have non-zero
winding, $g$ is independent of the Wilson line value $y_{B}$.
\subsubsection{Dirichlet boundary conditions}
The boundary entropy for the open string with Dirichlet boundary condition is
similary evaluated, starting again with the closed string channel.

 The boundary condition determining the boundary state is :
\begin{equation}
X|_{\tau=0}=y
\end{equation}
 leading to:
\begin{equation}
 w=0, \hspace{1cm} \alpha_{n}=\tilde{\alpha}_{-n}\ .
\end{equation}
From these conditions, for the boundary state located at the point $y$ we get
\begin{equation}
\label{DBSO}
|D_{y}\rangle=g_{D_{y}}\delta(x-y)\exp\left(\sum_{n>0}\frac{\alpha_{-n}\tilde{\alpha}_{-n}}{n}\right)
|0\rangle=g_{D_{y}}\sum_{p}e^{\frac{-ipy}
{R}}\exp\left(\sum_{n>0}\frac{\alpha_{-n}\tilde{\alpha}_{-n}}{n}\right)\left|\frac{p}{R},0\right
\rangle \ .
\end{equation}
Inserting this in (\ref{TC}) we have for the partition function in this channel:
\begin{eqnarray}
Z=g_{D_{y}}^{2}\langle D_{y}|\exp\left(\frac{-i\pi
p^{2}}{4R^{2}T}\right)\exp\left[\frac{-i\pi}{T}(N+\tilde{N}-\frac{1}{12})
\right]|D_{y}\rangle= \nonumber \\
\frac{g_{D_{y}}^{2}}{\eta(\tilde{q})}\sum_{p}\exp\left(\frac{-i\pi
p^{2}}{4R^{2}T}\right) 
=\frac{g_{D_{y}}^{2}}{\eta(\tilde{q})}\theta_{3}\left(-\frac{1}{4TR^{2}},0\right)\
.
\end{eqnarray}
 In order to analyse the open-string loop channel, 
  according to (\ref{LC}), the Hamiltonian must be expressed with the Dirichlet
boundary condition. Substituting the mode expansion of the coordinate $X$:
\begin{equation}
X=y+2wR\sigma+i\sum_{n \not=0}\frac{1}{n}\alpha_{n} \sin (n\sigma)\exp(-in\tau)
\end{equation}
 in the open-string Hamiltonian leads to:
\begin{equation}
H=2w^{2}R^{2}+N-\frac{1}{12}\ .
\end{equation}
 Finally, the partition function in this
channel is:
\begin{equation}
Z=\frac{1}{\eta(q)}\sum_{w}\exp\left(-4i\pi Tw^{2}R^{2}\right)= 
\frac{1}{\eta(q)}\theta_{3}(-4TR^{2},0)\ .
\end{equation}
 Equating the partition functions in the two channels and using (\ref{M}), one obtains:
\begin{equation}
\label{entd}
g_{D_{y}}^{2}=\frac{1}{2R}\ .
\end{equation}
Results (\ref{entn}) and (\ref{entd}) are consistent with $T$-duality.
The boundary entropy is independent of the position of the brane $y$.
Here it reflects the translation invariance of the disc diagramm.

\subsection{Boundary entropy on an orbifold}

In order to fully cover the $c=1$ moduli space, we must study the effect of
orbifolding on the boundary entropy. In general, the closed string moves
on an orbifold, as a result of modding out the target manifold by some discrete,
 finite symmetry group $G$, of order $n_{G}$. Into the original
unmodded theory, we introduce open strings satisfying some boundary conditions. If these
conditions are not invariant under the  $G$, then to maintain the symmetry
we must allow for all the different boundary conditions resulting from the
original ones by the action of  $G$. Generally, the original boundary conditions
 may be invariant under some subgroup $H$ of  $G$ of order  $n_{H}$.
 Then all the $n_{G}/n_{H}$ copies  resulting from applying  $G$ on these
conditions should also be allowed. In other words, we put into the original
 $G$-invariant target space a brane at a fixed point of  $H$ together with
all its  $n_{G}/n_{H}$ mirror images under  $G$, then identify all of them
by modding out by  $G$.

A closed-string boundary state $|B_{orb}\rangle$ on this orbifold brane is of the
form:
\begin{equation}
|B_{orb}\rangle=\sum_{h\in H} a_{h} \sum_{k\in G/H} |KB_{h}\rangle \ .
\end{equation}

Here, $B_{h}$ is a closed-string state on the brane twisted by the element
$h$ of the symmetry group preserving the brane, and the sum over $K$ runs
over all the mirror images, thus projecting down to a $G$-invariant state. 
The coefficients  $a_{h}$ are fixed by the theory. For $h=1$, $B_{1}$
 is the untwisted boundary state of the original unmodded theory. According
to the definition, the boundary entropy corresponding to this brane is:
\begin{equation}
\label{ORBENT}
g_{orb}=\langle0|B_{orb}\rangle=a_{1}\sum_{k\in G/H}\langle0|KB_{1}\rangle=(n_{G}/n_{H})a_{1}g\ ,
\end{equation}
$g$ being the boundary entropy of the unmodded model. In (\ref{ORBENT})
we used the fact that differently twisted states are mutually orthogonal and
that the closed string vacuum $|0\rangle$ is $G$-invariant.
To determine the number $a_{1}$ we use again the equality between the
two dual-channel representations of the annulus diagram whose two boundary
components are on the brane. The closed string channel gives: 
\begin{equation}
\label{PARTORB}
Z_{orb}=\langle B_{orb}|\tilde{q}^{H^{(P)}}|B_{orb}\rangle=\sum_{h\in H} |a_{h}|^{2}
\sum_{k,k' \in G/H}\langle KB_{h}|\tilde{q}^{H^{(P)}}|K'B_{h}\rangle \ ,
\end{equation}
$H^{(P)}$ being the closed string Hamiltonian. Since $H^{(P)}$ is $G$-invariant
\begin{equation}
\langle KB_{h}|\tilde{q}^{H^{(P)}}|K'B_{h}\rangle=
\langle B_{h}|\tilde{q}^{H^{(P)}}|K^{-1}K'B_{h}\rangle \ ,
\end{equation}
and (\ref{PARTORB}) becomes:
\begin{equation}
\label{PARTORB2}
Z_{orb}=n_{G}/n_{H}\sum_{h\in H} |a_{h}|^{2}\sum_{k\in
G/H}\langle B_{h}|\tilde{q}^{H^{(P)}}|KB_{h}\rangle \ .
\end{equation}
This should be equal to the open-string representation:
\begin{equation}
\label{PARTORB3}
Z_{orb}=\mathrm{Tr}\left(q^{H}P_{H}\right)=\sum_{k\in
G/H}\mathrm{tr}\left(q^{H_{k}}\frac{1}{n_{H}}\sum_{h\in H}h\right)\ .
\end{equation}
The ``Tr'' in (\ref{PARTORB3}) means summing over all states, including all
open-string twisted sectors, whereas the ``tr'' is the sum over states
in one, say $k$th, of them, where ${H_{k}}$ is the Hamiltonian of an open string connecting
the brane to its $K$ mirror. The time evolution operator $q^{H}$ is multiplied
by the projection onto the $H$ invariant space, $P_{H}=\frac{1}{n_{H}}\sum_{h\in
H}h$.
In fact each term in the sum over $K$ and $h$ in (\ref{PARTORB2}) and in (\ref{PARTORB3})
 represents a different annulus diagram, and the equality between the closed-string
representation,  (\ref{PARTORB2}), and the open-string description,
(\ref{PARTORB3}), should be valid term by term. In particular, for the term
corresponding to $h=1$, $K=1$, we know from the unmodded model that:
\begin{equation}
\langle B_{1}|\tilde{q}^{H^{(P)}}|B_{1}\rangle=\mathrm{tr}(q^{H_{1}})\ ;
\end{equation}
the coefficients of these terms in  (\ref{PARTORB2}) and
(\ref{PARTORB3}) should therefore be equal. This gives:
\begin{equation}
|a_{1}|^{2}=\frac{1}{n_{G}}\ .
\end{equation}
Equation (\ref{ORBENT}) then gives the relation between $g_{orb}$, the boundary
entropy on the orbifold, and $g$, that of the unmodded model:
\begin{equation}
\label{ORBENT2}
g_{orb}=\frac{n_{G}^{1/2}}{n_{H}}g \ .
\end{equation}
Specializing to the case of a $Z_{2}$ orbifold of a $c=1$ circle of the radius $R$,
we get for the Dirichlet brane at the fixed point, $n_{G}=n_{H}=2$, so that
(\ref{ORBENT2}) gives:
\begin{equation}
\label{EN2}
g_{orb}=\frac{g_{D}}{\sqrt{2}}=\frac{1}{2\sqrt{R}}\ .
\end{equation}
Similarly, for the Neumann boundary conditions with no variable Chan-Paton index
and (necessarily) no Wilson line,  $n_{G}=n_{H}=2$; hence:
\begin{equation}
\label{EN1}
g_{orb}=\frac{g_{N}}{\sqrt{2}}=\sqrt{\frac{R}{2}}\ .
\end{equation}
For two D-branes at points $x$ and $-x$ on the circle,  $n_{G}=2$ and
$n_{H}=1$:
\begin{equation}
g_{orb}=\sqrt{2}g_{D}=\frac{1}{\sqrt{R}}\ .
\end{equation}
For the $T$-dual case of the Neumann boundary conditions with the Chan-Paton
index taking two values and a non-zero Wilson line,  $n_{G}=2$ and
$n_{H}=1$:
\begin{equation}
g_{orb}=\sqrt{2}g_{N}=\sqrt{2R}\ .
\end{equation}
Let us note that, for the Neumann boundary condition, the entropy for the circle
(\ref{entn}) at $R=1/2\sqrt{2}$ is the same as the entropy
for orbifold (\ref{EN1}) at $R=1/\sqrt{2}$, as it should,
because these are exact values of $R$ at which the orbifold and
circle lines meet, in the closed-string moduli space.
For the Dirichlet boundary condition, the entropies for circle (\ref{entd}) and orbifold
(\ref{EN2}) coincide at the $T$-dual points: $\sqrt{2}$ for the circle, and
again $1/\sqrt{2}$ for the orbifold, because it is the self-dual point in this
case. \\
Similar considerations for the case of a $Z_{2}$ orbifold can be found 
in \cite{AO}.

\subsection{Ginsparg points}
At the self-dual radius $R=1/\sqrt{2}$ of the circle, the symmetry is enhanced
 to $SU(2)_{R}\times SU(2)_{L}$. Adding open strings may break this enhanced symmetry
 to some lower symmetry \cite{DL,GRGUT}. 
One can consider modding the open-string theory by  subgroups of this surviving symmetry.
In the Neumann case the surviving subgroup is the diagonal $SU(2)$ subgroup with generators:
\begin{equation}
J^{a}_{N}=\frac{1}{2}(J^{a}_{R}+J^{a}_{L})\ ;
\end{equation}
 in the Dirichlet case, the generators of the surviving subgroup are, by duality:
\begin{equation}
J^{a}_{D}=\frac{1}{2}(J^{a}_{R}-J^{a}_{L})\ ,
\end{equation}
where $J^{a}_{R}$ and $J^{a}_{L}$ are the generators of $SU(2)_{R}$ and
$SU(2)_{L}$ respectively.
This can be checked in the following way.
Consider for example the third component.
The generators of the enhanced $SU(2)\times SU(2)$ group can be realized as:
\begin{equation}
J^{3}_{R}(z)=-\frac{i}{\sqrt{2}}\partial x(z), \hspace{1cm} J^{\pm}_{R}=e^{\pm i\sqrt{2}x(z)},
\hspace{1cm} (z,J_{R} \rightarrow \bar{z},J_{L})\ .
\end{equation}
We see that the third component $J^{3}_{N}$ is just the momentum operator that annihilates
 the Neumann boundary state (\ref{NEU}).
 In the Dirichlet case $J^{3}_{D}$ is, by duality, a winding operator and, as such, annihilates
 the Dirichlet boundary state (\ref{DBSO}).
Modding out the $SU(2)$ symmetric theory, with either
Neumann or Dirichlet boundary  conditions (a different  $SU(2)$ for
each case), by some discrete subgroup of order $n$, we have in
eq. (\ref{ORBENT2})  $n_{G}=n_{H}=n$. Hence one has for the boundary entropy:
\begin{equation}
g_{orb}=\frac{g}{\sqrt{n}}\ .
\end{equation}

\subsection{Summary of the $c=1$ results}

 The following table is a collection of the $c=1$ results for the boundary entropy.
\vspace{1cm}

\begin{tabular}{||c|c|c||} \hline
Boundary condition  & Neumann & Dirichlet \\ \hline
Circle line & $\sqrt{R}$ & $\frac{1}{\sqrt{2R}}$ \\ \hline
Orbifold line &  &  \\ 
Generic points & $\sqrt{2R}$ & $\frac{1}{\sqrt{R}}$ \\ \cline{2-3}
Fixed Points & $\sqrt{\frac{R}{2}}$ & $\frac{1}{2\sqrt{R}}$ \\ \hline
Ginsparg points & & \\
Tetrahedral & $2^{-1/4}/\sqrt{12}$ & $2^{-1/4}/\sqrt{12}$ \\  \cline{2-3}
Octahedral &  $2^{-1/4}/\sqrt{24}$ & $2^{-1/4}/\sqrt{24}$ \\  \cline{2-3}
Icosahedral &   $2^{-1/4}/\sqrt{60}$ & $2^{-1/4}/\sqrt{60}$ \\ \hline
\end{tabular}

\vspace{1cm}

We can  find, at every point of the moduli space, the boundary condition providing
the least value of the entropy. This boundary condition will be preferable 
in the sense of boundary renormalization group.
From the table, we see that for $R>2^{-1/2}$ the
entropy is smaller for Dirichlet boundary condition, and for $R<2^{-1/2}$ it is
smaller for Neumann.
At the self-dual point and at the Ginsparg points, entropy is the same for both types of boundary
conditions. All these results are visualized in fig. 1.
The more stable (Neumann or Dirichlet) boundary condition at each region in the
moduli space is indicated.
The orbifold line is depicted here only for the case of the fixed point.

\begin{figure}[h]
\begin{picture}(500,200)(-50,-50)
\put(0,0){\circle*{5}}
\put(0,0){\line(80,0){350}}
\put(80,0){\line(0,-100){100}}
\put(160,0){\circle*{5}}
\put(250,0){\line(0,100){100}}
\put(80,-100){\circle*{5}}
\put(40,5){N}
\put(85,-50){N}
\put(120,5){N}
\put(200,5){D}
\put(255,50){D}
\put(300,-12){D}
\put(355,0){$R_{c}$}
\put(250,105){$R_{o}$}
\put(150,-15){$1/\sqrt{2}$}
\put(245,-15){$\sqrt{2}$}
\put(67,8){$1/2\sqrt{2}$}
\put(45,-20){$1/\sqrt{2}$}
\put(255,8){$1/\sqrt{2}$}
\put(160,50){$\bullet$}
\put(170,50){I}
\put(160,70){$\bullet$}
\put(170,70){O}
\put(160,90){$\bullet$}
\put(170,90){T}
\put(0,-12){0}
\put(70,-100){0}
\end{picture}
\vspace{2cm}
\caption{Map of the preferred boundary conditions in the $c=1$ moduli space}
\end{figure}
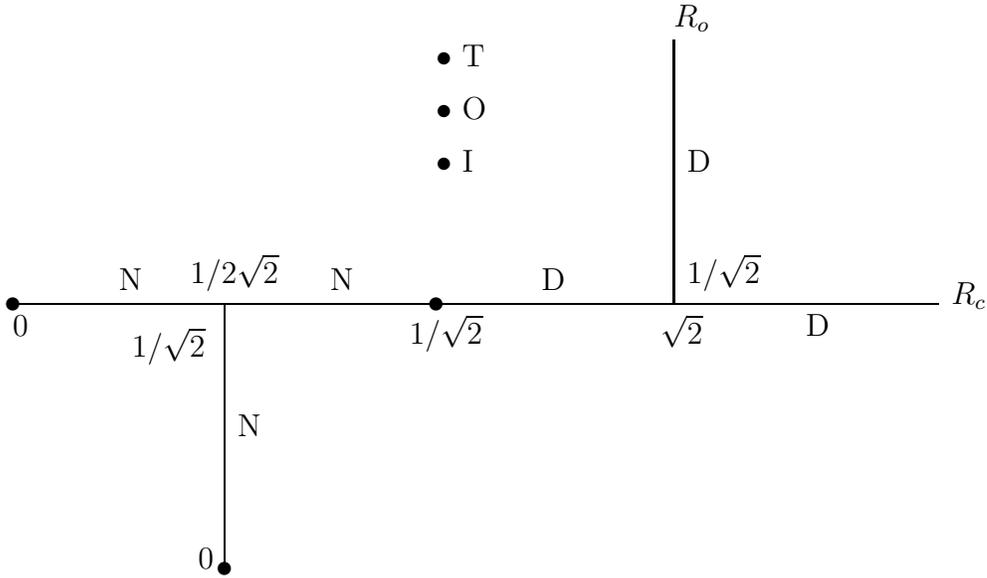
\vspace{1cm}
The fact that $g_{D}\leq g_{N}$ for $R$ greater than the radius of the self-dual
 point, is in accordance with its expected role as a stability parameter.
Note that the self-dual radius for the general value of the string slope
$\alpha'$ self-dual radius equals $\sqrt{\alpha'}$. 
 If the  $c=1$ $X$-coordinate is
compactified on a circle of radius $R$ in one of the additional transverse
$d$ spatial coordinates with the Neumann boundary condition, the Dirichlet boundary
condition on $X$ corresponds to a $d$-brane while the Neumann boundary condition
describes a ($d+1$)-brane wrapped on the circle of radius $R$. The tension of a
($d+1$)-brane is related to that of the $d$-brane by $T_{d+1}=T_{d}/2\pi\sqrt{\alpha'}$ \cite{POLCH}.
The energy density in the transverse world is $T_{d}$ for the Dirichlet
boundary condition, and $2\pi RT_{d+1}= T_{d}R/\sqrt{\alpha'}$ for the Neumann
boundary condition on $X$. Thus for $R$ larger than $\sqrt{\alpha'}$,
the Dirichlet boundary condition on $X$ gives less energy density in the transverse
world than the Neumann boundary condition, and the Neumann vacuum may decay to the
Dirichlet vacuum. For $R$ smaller than  $\sqrt{\alpha'}$, the Dirichlet boundary
condition requires higher energy density than  the Neumann boundary condition, and
the Dirichlet vacuum could be unstable, in accordance with its
higher value of $g$. 

\section{Computation of the boundary entropy for Neumann and Dirichlet boundary conditions for the
bosonic string compactified on a torus}
\subsection{Description of a bosonic string compactified on a torus} 
The world-sheet action describing the bosonic string moving in a toroidal
background is \cite{NSW}:
\begin{equation}
\label{ACT}  
S  =  \frac{1}{2\pi}\int_{0}^{\pi}d\sigma\int d\tau
G_{ij}\partial_{\alpha}X^{i}\partial^{\alpha}X^{j} 
+   \varepsilon^{\alpha\beta}B_{ij}\partial_{\alpha}X^{i}\partial_{\beta}X^{j}\
.
\end{equation} 
Here $X^{i}$ are dimensionless coordinates whose periodicities are chosen to be
$2\pi$, namely
\mbox{\( X^{i}\approx X^{i}+2\pi m^{i} \)};
 $E$ is the matrix whose symmetric part is $G$ and the antisymmetric part is $B$: \mbox{$E=G+B$}.
The canonical momentum $P_{i}$ associated with  $X^{i}$ is given by
\begin{equation}
\label{MOM}  
\pi P_{i}=G_{ij}\partial_{\tau}X^{j}+B_{ij}\partial_{\sigma}X^{j}\ ;
\end{equation}
$X^{i}$ is given by : 
\begin{equation}
\label{MC}
X^{i}(\sigma ,\tau )  =  x^{i}+2w^{i}\sigma+ \tau G^{ij}(p_{j}-2B_{jk}w^{k}) 
 + \frac{i}{2} \sum_{n \not= 0}\frac{1}{n} [\alpha_{n}^{i}(E)e^{-2in(\tau -\sigma)}+
\tilde{\alpha}_{n}^{i}(E)e^{-2in(\tau +\sigma)}],
\end{equation}
$P_{i}$ is given by:
\begin{equation}
\label{MM}   
\pi P_{i}( \sigma , \tau )= p_{i}+\frac{i}{2}\sum_{n \not=0 } [ E_{ij}^{t}\alpha_{n}^{j}(E)
e^{-2in(\tau -\sigma)}+E_{ij}\tilde{\alpha}_{n}^{j}(E)e^{-2in(\tau +\sigma)} ]\
.
\end{equation}
The coordinates $X^{i}$ are split into left and right modes:
\begin{eqnarray}
X^{i}_{L}=x^{i}/2+(w^{i}+G^{ik}p_{k}/2-G^{ij}B_{jk}w^{k})(\tau+\sigma)+\mathrm{oscillators}  \nonumber \\
X^{i}_{R}=x^{i}/2+(-w^{i}+G^{ik}p_{k}/2-G^{ij}B_{jk}w^{k})(\tau -
\sigma)+\mathrm{oscillators}\ ,
\end{eqnarray}
Whence:
\begin{eqnarray}
P^{i}_{L}=w^{i}+G^{ik}p_{k}/2-G^{ij}B_{jk}w^{k}+\mathrm{oscillators} \nonumber \\  
P^{i}_{R}=-w^{i}+G^{ik}p_{k}/2-G^{ij}B_{jk}w^{k}+\mathrm{oscillators}
\end{eqnarray}
\begin{equation}
H=\frac{1}{2}(P_{L}^{2}+P_{R}^{2})
\end{equation}
\begin{equation}
\label{MH}  
H  =  \frac{1}{4}[p_{i}(G^{-1})^{ij}p_{j}+4w^{i}(G-BG^{-1}B)_{ij}w^{j}+4w^{i}B_{ik}(G^{-1})^{kj}p_{j}]+N+\tilde{N}, 
\end{equation}
where $N$, $\tilde{N}$ are number operators: 
\begin{equation}
N=\sum_{n>0}\alpha_{-n}^{i}(E)G_{ij}\alpha_{n}^{j}(E),\,
\tilde{N}=\sum_{n>0}\tilde{\alpha}_{-n}^{i}(E)G_{ij}\tilde{\alpha}_{n}^{j}(E)
\end{equation}
and $\alpha$, $\tilde{\alpha}$ satisfy the commutation relation:
\begin{equation}
\label{CR}
[\alpha_{n}^{i}(E),\alpha_{m}^{i}(E)]=[\tilde{\alpha}_{n}^{i}(E),\tilde{\alpha}_{m}^{i}(E)]=m(G^{-1})^{ij}\delta_{m+n,0}\
.
\end{equation}
For the open-string case, there are few 
changes: a power of $2$ does not appear in the exponent, the left and right
modes are no longer independent, but constrained by the
boundary conditions.
In the closed-string channel, $w^{i}$ and $p_{i}$ are both integer-valued, whereas
for the open string only one of them is an integer, depending on the boundary
condition.
 
The possible boundary conditions follow from variation of the action (\ref{ACT}):
\begin{equation}
\delta S=-\frac{1}{\pi}\int \int_{D} G_{ij}\Box X^{j}\delta X^{i}+ 
\frac{1}{\pi}\int_{\partial
D}(G_{ij}\partial_{n}X^{j}+\varepsilon^{n\beta}B_{ij}\partial_{\beta}X^{j})\delta
X^{i}\ ,
\end{equation} 
where $D$ is the range of integration and $n$ denotes the normal to the boundary.
This leads to two kinds of boundary conditions: it is either
\(G_{ij}\partial_{n}X^{j}+\varepsilon^{n\beta}B_{ij}\partial_{\beta}X^{j}=0 \),
the Neumann boundary condition, 
or \( \delta X^{j}=0 \), the Dirichlet boundary condition.
 
\subsection{ Neumann boundary conditions} 

The boundary entropy is again obtained by equating the partition function calculated in the
 closed- and open-string channel, respectively.

\subsubsection{The closed-string tree channel}

The boundary  state $|N\rangle$ is calculated by solving the various boundary
constraints imposed by the boundary conditions.
In the closed-string channel the boundary is located at $\tau=0$, and therefore:
\begin{equation} 
G_{ij}\partial_{\tau}X^{j}+B_{ij}\partial_{\sigma}X^{j}=0\ .
\end{equation}
From (\ref{MOM}) we see that it is just a condition of vanishing momentum flow
at the boundary, which should be imposed on the boundary state:
\begin{equation}
\label{van}
P_{i}|N\rangle=0 \ .
\end{equation}
From the mode expansion for the momenta (\ref{MM}), we obtain:
\begin{equation}
\label{ZV}
p_{i}=0 
\end{equation}
for the zero-mode part and the constraint:
\begin{equation}
\label{VO}
E_{ij}^{t}\alpha_{n}^{j}=- E_{ij}\tilde{\alpha}_{-n}^{j} 
\end{equation}
 for the left- and right-handed oscillator modes.
The oscillator part of $|N\rangle$ is obtained by rewriting (\ref{VO}) in matrix form
and solving for $\alpha_{n}$, to
get:
\begin{equation}
\label{AO}
\alpha_{n}=-(E^{t})^{-1}E\tilde{\alpha}_{-n},
\end{equation}
From this, $|N\rangle$ is found to be:
\begin{equation}
\label{NBS1}
|N\rangle=g_{N}\sum_{w}
\exp\left(\sum_{n>0}-\frac{\alpha_{-n}^{t}G(E^{t})^{-1}E\tilde{\alpha}_{-n}}{n}\right)|0,w\rangle
\ .
\end{equation}
Finally, putting together (\ref{NBS1}) and the formula for the Hamiltonian
 (\ref{MH}) in (\ref{TC}),
one arrives  at the following expression for the partition function:
\begin{eqnarray}
\label{PFC}
Z=\langle N|\exp(-\pi iH^{(P)}/T)|N\rangle
  =
  g_{N}^{2}\frac{\sum_{w}e^{-i\pi\frac{1}{T}w^{t}(G-BG^{-1}B)w}}{\eta^{d}(\tilde{q})}
 \nonumber \\ 
   =  g_{N}^{2}\frac{\Theta(-(G-BG^{-1}B)/T,0)}{\eta^{d}(\tilde{q})} \ ,
\end{eqnarray}
where
\(\Theta(\Omega,\vec{z})=\sum_{\vec{n}\epsilon Z^{g}} \exp(\pi i
\vec{n}^{t}\Omega\vec{n}+
2\pi i\vec{n}^{t}\vec{z}) \)
is the theta function with the modular matrix $\Omega$ .

\subsubsection{The open-string loop channel}

In this channel, according to (\ref{LC}) we should compute the trace of the
 matrix density for the Hamiltonian $H$
corresponding to the Neumann boundary conditions. 
Here  the boundary is located at $\sigma=0$,  and therefore:
\begin{equation}
\label{NB}
G_{ij}\partial_{\sigma}X^{j}+B_{ij}\partial_{\tau}X^{j}=0\ .
\end{equation}
Substituting the mode expansion of $X^{i}$  (\ref{MC}) 
 one gets the following constraint for the values
of $p_{i}$ and $w^{i}$ in the zero-mode part:
\begin{equation}
\label{WN}
2(G-BG^{-1}B)w=-BG^{-1}p \Longrightarrow w=\frac{1}{2}(B-GB^{-1}G)^{-1}p\ .
\end{equation} 
Note that, in the open-string channel and in the presence of  Neumann boundary
 conditions, the ``winding'' numbers need not be integers. 
 This is unlike the situation in the closed-string channel.
The momenta $p_{i}$ remain integer-valued.
Substituting the  value of $w$ in the Hamiltonian, we find:
\begin{equation}
H=\frac{1}{2}p^{t}(G-BG^{-1}B)^{-1}p+N \ ,
\end{equation} 
leading to:
\begin{equation}
\label{PFO}
Z=\frac{\sum_{p}e^{-i\pi Tp^{t}(G-BG^{-1}B)^{-1}p}}{\eta^{d}(q)} 
 =  \frac{\Theta(-(G-BG^{-1}B)^{-1}T),0)}{\eta^{d}(q)}\ .
\end{equation}
Recalling the following modular transformation properties of $\Theta$ and
 $\eta$ functions ~\cite{DM}:
\begin{eqnarray}
\label{MTP}
\Theta(\Omega^{-1},0)=(\mathrm{det}\Omega)^{1/2}\Theta(\Omega,0) \nonumber  \\
\eta(q(-1/T))=(-T)^{1/2}\eta(q(-T)) \ ,
\end{eqnarray}
equating (\ref{PFC}) and (\ref{PFO}), and using (\ref{MTP}), we get:
 
\begin{equation}
\label{NE}
g_{N}=(\mathrm{det}(G-BG^{-1}B))^{1/4}=\frac{(\mathrm{det}E)^{1/2}}{(\mathrm{det}G)^{1/4}}\
.
\end{equation} 
\subsection{Dirichlet boundary conditions}

\subsubsection{The closed-string tree channel}
 
In this channel the boundary is located at $\tau=0$, and we should therefore impose the Dirichlet boundary condition
in the form:
\begin{equation}
\delta X^{i}|_{\tau=0}=0\ .
\end{equation}
Substituting here $X^{i}$ from (\ref{MC}) we get:
\begin{eqnarray}
w^{i}=0 \nonumber  \\
\alpha_{n}^{i}=\tilde{\alpha}_{-n}^{i}\ .
\end{eqnarray}
Following the same reasoning as in the Neumann case, we find the boundary
states $|D\rangle$ to be:
\begin{equation}
\label{DBS}
|D\rangle=g_{D}\sum_{p}
\exp\left(\sum_{n>0}\frac{\alpha_{-n}^{t}G\tilde{\alpha}_{-n}}{n}\right)|p,0\rangle
\ .
\end{equation}
Substituting (\ref{DBS}) in the formula for the partition function in the
closed-string channel (\ref{TC}),
and using for the Hamiltonian (\ref{MH}), we obtain:
\begin{equation}
\label{PFCD}
Z=\langle D|\exp(-\pi iH^{(P)}/T)|D\rangle
=g_{D}^{2}\frac{\sum_{p}e^{-i\pi\frac{1}{4T}p^{t}G^{-1}p}}{\eta^{d}(\tilde{q})}  
=g_{D}^{2}\frac{\Theta(-G^{-1}/4T,0)}{\eta^{d}(\tilde{q})} \ .
\end{equation}

\subsubsection {The open-string loop channel}

In this channel the boundary is located at  $\sigma=0$ and the  Dirichlet
 boundary condition is of the form:
\begin{equation}
\label{DX}
\delta X^{i}|_{\sigma=0}=0\ .
\end{equation}
Inserting $X^{i}$ from (\ref{MC}) to (\ref{DX}), we obtain the following
 constraint for the values of $w^{i}$ and $p_{i}$:
\begin{equation}
\label{DBC}
p_{j}=2B_{jk}w^{k}\ .
\end{equation}
In the Dirichlet case the winding numbers are integer-valued while $p_{i}$, as
 determined by (\ref{DBC}),
are not necessarily integers. This also follows from $T$-duality.
Substituting (\ref{DBC}) in the Hamiltonian gives:
\begin{equation}
\label{DH}
H=2w^{i}G_{ij}w^{j}+N
\end{equation}
and :
\begin{equation}
\label{PFOD}
Z=\frac{\sum_{w}e^{-4i\pi Tw^{t}Gw}}{\eta^{d}(q)} 
 =  \frac{\Theta(-4GT,0)}{\eta^{d}(q)}\ .
\end{equation}
Equating (\ref{PFCD}) and (\ref{PFOD}), and using the modular transformation
 properties (\ref{MTP}), we 
extract the factor $g_{D}$:
\begin{equation}
\label{DE}
g_{D}=\frac{1}{(\mathrm{det}4G)^{1/4}}=\frac{1}{2^{d/2}(\mathrm{det}G)^{1/4}}\ .
\end{equation}
We see that, in the case of the Dirichlet boundary condition, the  partition function does not
depend on the topological $B$-term. The explanation is as follows:
for Dirichlet boundary conditions, the world-sheet disc has its boundary shrunk
to a point, and it is therefore mapped into a sphere in target space.
We may conclude that 
the map of the world-sheet into the target-space torus is an element of the
$\pi_{2}(T)$ group, which
is  known to be trivial. Hence, in target space any image of the world-sheet disc, that
contributes to the path integral is a two-sphere, which is necessarily the boundary
of some three-manifold. On such a sphere the integral of the closed two-form $B$
vanishes and therefore the disc diagram is $B$-independent.

\section{ T-duality relations between the Neumann and Dirichlet boundary conditions}
We next show that there exists a particular element of the $T$-duality  group
 $O(d,d,Z)$ mapping all results for the Neumann condition
to the Dirichlet one, namely:
\begin{equation}
E\rightarrow \frac{1}{4E}\ .
\end{equation}
Under the action of this element, the following transformations occur \cite{GPR}:
\begin{equation}
\label{DTP1}
w\leftrightarrow p
\end{equation}
\begin{equation}
\label{DTP2} 
G\rightarrow G'=\frac{1}{4}(G-BG^{-1}B)^{-1}=\frac{1}{4}(E^{t})^{-1}GE^{-1}=\frac{1}{4}E^{-1}G(E^{t})^{-1}
\end{equation}
\begin{equation}
\label{DTP3}
B\rightarrow B'=\frac{1}{4}(B-GB^{-1}G)^{-1},\hspace{1cm} G^{-1}B\rightarrow -BG^{-1} 
\end{equation}
\begin{equation}
\label{DTP4}
\tilde{\alpha}_{n}(E) \rightarrow \tilde{\alpha}_{n}(E')=2E\tilde{\alpha}_{n}(E),\hspace{1cm}
\alpha_{n}(E) \rightarrow  \alpha_{n}(E')=-2E^{t}\alpha_{n}(E)\ .
\end{equation}
Now we want to check that, under this transformation,
\begin{enumerate}
\item Neumann boundary condition $\longrightarrow$ Dirichlet  boundary condition
\item $g_{N}$ Neumann boundary entropy $\longrightarrow$ $g_{D}$  Dirichlet boundary entropy
\item  Neumann boundary state $|N\rangle$ $\longrightarrow$ Dirichlet boundary
state $|D\rangle$ \ .
\end{enumerate}
The first statement follows from the comparison of formulas (\ref{WN}) and (\ref{DBC}), taking
into account (\ref{DTP1}) and  (\ref{DTP3}).
The second statement follows from formulas (\ref{NE}), (\ref{DE}) and (\ref{DTP2}).
As for  the third statement,
from (\ref{NBS1}) and (\ref{DBS}):
\begin{eqnarray}
|N\rangle=g_{N}\sum_{w} \exp\left(\sum_{n>0}-\frac{\alpha_{-n}^{t}G(E^{t})^{-1}E\tilde{\alpha}_{-n}}{n}\right)|0,w\rangle \nonumber \\
|D\rangle=g_{D}\sum_{p}
\exp\left(\sum_{n>0}\frac{\alpha_{-n}^{t}G\tilde{\alpha}_{-n}}{n}\right)|p,0\rangle\
.
\end{eqnarray}
We have already noted that $g_{N}\rightarrow g_{D}$ and $w\rightarrow p$, and it is left to check that:
\begin{equation}
-\alpha_{-n}^{t}G(E^{t})^{-1}E\tilde{\alpha}_{-n}\rightarrow\alpha_{-n}^{t}G\tilde{\alpha}_{-n}\
.
\end{equation}
Using now (\ref{DTP2}) and (\ref{DTP4}), we verify that this indeed occurs:
\begin{equation}
-\alpha_{-n}^{t}G(E^{t})^{-1}E\tilde{\alpha}_{-n}\rightarrow 
 -(-2E^{t}\alpha_{n}(E))^{t}\left(\frac{1}{4}E^{-1}G(E^{t})^{-1}\right)(4E^{t})\left(\frac{1}{4}E^{-1}\right)2E\tilde{\alpha}_{n}(E)
=\alpha_{-n}^{t}G\tilde{\alpha}_{-n}\ .
\end{equation}

\section{ Boundary entropy for the general case with an arbitrary number of Dirichlet
 and Neumann coordinates}
As a more general case we require the first $k$ coordinates to satisfy the
Dirichlet boundary conditions, and the last $(d-k)$ coordinates to satisfy
 the Neumann ones.
The boundary conditions are now expressed in the following form:
\begin{equation}
\label{DK}
\delta X^{r}|_{\tau=0}=0 \Rightarrow w^{r}=0, \hspace{1cm}
\alpha_{n}^{r}=\tilde{\alpha}_{-n}^{r},  \hspace{0.5cm} r=1, \ldots ,k\ ,
\end{equation}
\begin{equation}
\label{NK}
P_{\alpha}=0 \Rightarrow p_{\alpha}=0, \hspace{1cm}  E_{\alpha j}^{t}\alpha_{n}^{j}+E_{\alpha
j}\tilde{\alpha}_{-n}^{j}=0, \hspace{0.5cm} \alpha=k+1, \ldots ,d\ .
\end{equation}
Substituting the  values of $w$ and $p$ found in (\ref{DK}) and (\ref{NK}) in (\ref{MH}),
we get for zero-modes Hamiltonian:
\begin{equation}
H=L^{t}\Omega L
\end{equation}
where $L$ is the vector: \[ \left( \begin{array}{c} p_{r} \\ w^{\alpha} \end{array} \right) \] and $\Omega$ is the matrix:
\begin{equation}
\left( \begin{array}{cc} 
(G^{-1})^{rs}/4 & B_{\alpha i}G^{ir}/2 \\
B_{\alpha i}(G^{-1})^{is}/2 & G_{\alpha \beta}-B_{\alpha
i}(G^{-1})^{ij}B_{j\beta} \end{array} \right)\ .
\end{equation}
(As was mentioned, we mean here by $i,j$ sums over all coordinates from $1$ to
$d$ ).
Following previous considerations, we find that the square of the entropy is now equal to
the square root of the determinant of $\Omega$:
\begin{equation}
g_{k}^{2}=\sqrt{\mathrm{det}\Omega}\ .
\end{equation}
Let us analyse various properties of this expression.
First of all, let us note that by multiplying each $r$th row from the first $k$ rows by
$2B_{r\alpha}$, adding all of them to the $\alpha$th row, and repeating the same procedure
 with the first $k$ columns, we can completely eliminate all
components of $B$ containing at least one index belonging to the Dirichlet set:
\begin{equation}
\label{ENTR}
g_{k}^{2}=\sqrt{\mathrm{det}\Omega'}\ ,
\end{equation}
where $\Omega'$ is: 
\begin{equation}
\Omega'=\left( \begin{array}{cc} 
(G^{-1})^{rs}/4 & B_{\alpha \gamma}G^{\gamma r}/2 \\
B_{\alpha \gamma}(G^{-1})^{\gamma s}/2 & G_{\alpha \beta}-B_{\alpha
\delta}(G^{-1})^{\delta\gamma}B_{\gamma\beta}
 \end{array} \right)\ .
\end{equation}
We see that the entropy, or the vacuum degeneracy of the theory, depends only on components of $B$
tangent to the corresponding D-brane.
It is possible to explain this along the same lines as  used for explaining the independence
of the partition function on some $B$-terms in the pure Dirichlet case.

$g_{k}$ is given by a path integral over disc diagrams embedded in the $d$
torus with the boundary on the $d-k$ dimensional D-brane. For any given
embedded disc $D$ in this path integrand, the boundary is some closed curve in
the brane. Being the boundary of a  disc $D$, this curve  does not  wrap any
non-trivial cycle of the torus; therefore, it is also the boundary of some other
disc, $D'$, which lies entirely inside the  $(d-k)$-brane. The union of  $D$ and  
 $D'$ is then topologically a sphere embedded in the  $d$-torus.
The argument in the previous section implies that:
\begin{equation}
\int_{D}B+\int_{D'}B=0\ .
\end{equation}
The $B$ dependence of every embedding in the path integral can be expressed
in terms of a $B$ integral over a disc  $D'$ lying entirely in the brane.
This integral depends only on the component of $B$ tangent to the brane.

Let us check also here that all $g_{k}$ can be derived
 from one another by means of the product of  transformations of the factorized
dualities \cite{GPR}:
\begin{equation}
D_{i}=\left( \begin{array}{cc}
I-e_{i} & e_{i}/2 \\
 e_{i}/2 & I-e_{i} \end{array} \right)\ ,
\end{equation}
where $I$ is a $d$-dimensional identity matrix, and $e_{i}$ is zero, except for
the $ii$ component, which is $1$.
Let us check how the $g_{N}$ that was found earlier transforms to $g_{1}$ under the $D_{1}$
transformation; the generalization to other cases will be straightforward.
By formula (2.49) in \cite{GPR}, we deduce that under any \(  \left( \begin{array}{cc}
a & b \\
c & d  \end{array} \right) \) element of the $O(d,d,Z)$ group the matrix $G-BG^{-1}B$ transforms
as follows:
\begin{equation}
G-BG^{-1}B \rightarrow
a(G-BG^{-1}B)a^{t}+a(BG^{-1})b^{t}-b(G^{-1}B)a^{t}+bG^{-1}b^{t}\ .
\end{equation}
Noting that the right or left multiplication by any matrix on $e_{1}$ amounts to
setting to zero all elements except the first row or column respectively, and right
or left multiplication by $I-e_{1}$ reduces to setting to zero elements of the first
row or column respectively, we arrive at the mentioned result.

We turn to the  analysis of the preferred type of 
boundary conditions at some interesting regions of the moduli space of toroidal
compactification. Namely we want to find at
every point in moduli space what type
of boundary conditions provides the least value of the boundary entropy, and
consequently which is preferable in the sense of the boundary renormalization
group.

To begin with, we consider the boundary entropy for the case of a diagonal
metric and a vanishing background $B$-field.
From (\ref{ENTR}) we see that here the squared entropy generally has the form:
\begin{equation}
g_{k}^{2}=\frac{R_{k+1}\cdots R_{d}}{2^{k}R_{1}\cdots R_{k}}\ ,
\end{equation}
where $R_{i}$ is the radius of the dimension $X_{i}$.
We see that the type of the preferred boundary condition gets changed as the radius of some
coordinate reaches the self-dual value.
Finally to see the influence of the background $B_{12}$ on the value of the
boundary entropy we consider the case of $c=2$. 
For $c=2$ we have three types of boundary conditions: 
\begin{enumerate}
\item
both coordinates satisfy Neumann conditions,
\item
one, say the first, satisfies the Dirichlet boundary conditions, and the second
the Neumann boundary conditions,
\item
both satisfy the Dirichlet boundary conditions.
\end{enumerate}
From (\ref{ENTR}) we have the following expressions of entropy corresponding to these cases:
\begin{equation}
g_{NN}^{2}=\frac{G_{11}G_{22}-G_{12}^{2}+B_{12}^{2}}{\sqrt{\mathrm{det}G}} \hspace{2cm}
g_{DN}^{2}=\frac{G_{22}}{4\sqrt{\mathrm{det}G}} \hspace{2cm}
g_{DD}^{2}=\frac{1}{4\sqrt{\mathrm{det}G}}\ .
\end{equation}
For further analysis it is convenient to organize the four real data set of moduli
space $G_{11}$, $G_{22}$, $G_{12}$, $B_{12}$ into two complex coordinates $\rho$
and $\tau$, in the following manner \cite{GPR}:
\begin{equation}
\tau\equiv\tau_{1}+i\tau_{2}=\frac{G_{12}}{G_{22}}+\frac{i\sqrt{\mathrm{det}G}}{G_{22}},
\hspace{2cm}
\rho\equiv\rho_{1}+i\rho_{2}=B_{12}+i\sqrt{\mathrm{det}G}\ .
\end{equation}
In these coordinates the  expressions for the entropies take the forms:
\begin{equation}
g_{NN}^{2}=\frac{\rho_{1}^{2}+\rho_{2}^{2}}{\rho_{2}} \hspace{2cm}
g_{DN}^{2}=\frac{1}{4\tau_{2}} \hspace{2cm}
g_{DD}^{2}=\frac{1}{4\rho_{2}}\ .
\end{equation}
It is interesting to note that in these coordinates all expressions depend only
on three parameters $\rho_{1}$, $\rho_{2}$, $\tau_{2}$ and are independent of
$\tau_{1}$.
Comparing values of the entropy for different boundary conditions gives us the
following:
\begin{equation}
\label{UNEQ1}
g_{NN}\geq g_{DD} \hspace{0.5cm} \mathrm{if} \hspace{0.5cm}
\rho_{1}^{2}+\rho_{2}^{2}\geq\frac{1}{4} 
\end{equation}
\begin{equation}
\label{UNEQ2}
g_{NN}\geq g_{DN}  \hspace{0.5cm} \mathrm{if} \hspace{0.5cm}
\rho_{1}^{2}+\rho_{2}^{2}\geq\frac{\rho_{2}}{4\tau_{2}} 
\end{equation}
\begin{equation}
\label{UNEQ3}
g_{DN}\geq g_{DD}  \hspace{0.5cm} \mathrm{if} \hspace{0.5cm}
\rho_{2}\geq\tau_{2}\ .
\end{equation}
In order to visualize  these inequalities, consider  
two cases separately. First  when $\rho_{1}^{2}+\rho_{2}^{2}\geq 1/4$, which means that
we consider a point in the moduli space out of the cylinder $\rho_{1}^{2}+\rho_{2}^{2}=1/4$,  and  the second when
 $\rho_{1}^{2}+\rho_{2}^{2}\leq 1/4$, corresponding to a point inside the cylinder.

In the first case, $g_{NN}\geq g_{DD}$ and in order to find the least value we
should compare  $g_{DD}$ with $g_{DN}$. From (\ref{UNEQ3}) we see that in this
region over the hyperplane $\rho_{2}=\tau_{2}$ the least value is provided by  $g_{DN}$
and for the region below by  $g_{DD}$.

In the second region, when the point is inside the cylinder,  $g_{NN}\leq g_{DD}$
and we should compare  $g_{NN}$ with  $g_{DN}$. From (\ref{UNEQ2}) we see that in
this region over the hypersurface $\rho_{1}^{2}+\rho_{2}^{2}=\rho_{2}/4\tau_{2}$
 the least value of the entropy is provided by  $g_{DN}$ and for the region
below it  by $g_{NN}$.
Graphically all these results are depicted in  fig. 2.  
\begin{figure}[ht]
\epsfig{file=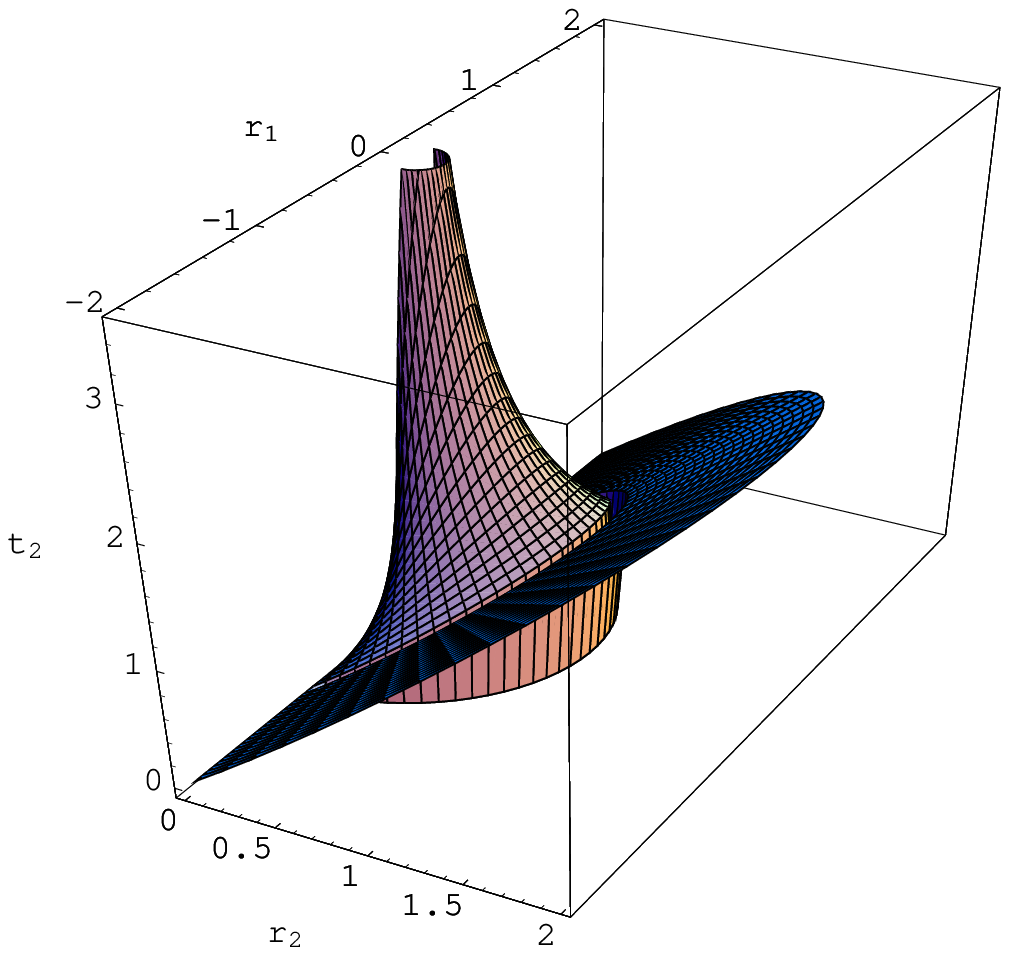,height=160mm,width=135mm}
\caption{Map of the prefered boundary conditions in $c=2$ moduli space. 
The hyper plane $\rho_{2}=\tau_{2}$ is denoted in black. The grey figure
below, marked by the vertical lines, is the cylinder
$\rho_{1}^{2}+\rho_{2}^{2}=1/4$. The curved surface above is the hypersurface
$\rho_{1}^{2}+\rho_{2}^{2}=\rho_{2}/4\tau_{2}$. In the region between the hypersurface
and the hyperplane the least value for the entropy is provided by $g_{DN}$, between the cylinder
and the hyperplane by $g_{DD}$, and inside of the cylinder and below the
hypersurface by $g_{NN}$.}
\end{figure}
\section{Acknowledgements}
We thank Christoph Schweigert for discussions.
 The work of E. R. is partially
supported by the Israel Academy of Sciences and Humanities--Centers
of Excellence Programme, and the American-Israel Bi-National Science
Foundation.  The work of S.E. is supported by the Israel Academy of Sciences 
and Humanities--Centers of Excellence Programme.

\end{document}